\PassOptionsToPackage{unicode}{hyperref}
\PassOptionsToPackage{hyphens}{url}
\documentclass[
]{article}
\usepackage{xcolor}
\usepackage[margin=1in]{geometry}
\usepackage{amsmath,amssymb}
\usepackage{iftex}
\ifPDFTeX
  \usepackage[T1]{fontenc}
  \usepackage[utf8]{inputenc}
  \usepackage{textcomp} 
\else 
  \usepackage{unicode-math} 
  \defaultfontfeatures{Scale=MatchLowercase}
  \defaultfontfeatures[\rmfamily]{Ligatures=TeX,Scale=1}
\fi
\usepackage{lmodern}
\ifPDFTeX\else
\fi
\IfFileExists{upquote.sty}{\usepackage{upquote}}{}
\IfFileExists{microtype.sty}{
  \usepackage[]{microtype}
  \UseMicrotypeSet[protrusion]{basicmath} 
}{}
\makeatletter
\@ifundefined{KOMAClassName}{
  \IfFileExists{parskip.sty}{%
    \usepackage{parskip}
  }{
    \setlength{\parindent}{0pt}
    \setlength{\parskip}{6pt plus 2pt minus 1pt}}
}{
  \KOMAoptions{parskip=half}}
\makeatother
\usepackage{graphicx}
\makeatletter
\newsavebox\pandoc@box
\newcommand*\pandocbounded[1]{
  \sbox\pandoc@box{#1}%
  \Gscale@div\@tempa{\textheight}{\dimexpr\ht\pandoc@box+\dp\pandoc@box\relax}%
  \Gscale@div\@tempb{\linewidth}{\wd\pandoc@box}%
  \ifdim\@tempb\p@<\@tempa\p@\let\@tempa\@tempb\fi
  \ifdim\@tempa\p@<\p@\scalebox{\@tempa}{\usebox\pandoc@box}%
  \else\usebox{\pandoc@box}%
  \fi%
}
\def\fps@figure{htbp}
\makeatother
\NewDocumentCommand\citeproctext{}{}

\makeatletter
 \let\@cite@ofmt\@firstofone
 \def\@biblabel#1{}
 \def\@cite#1#2{{#1\if@tempswa , #2\fi}}
\makeatother
\newlength{\cslhangindent}
\newlength{\csllabelwidth}
\newenvironment{CSLReferences}[2] 
 {\begin{list}{}{%
  \setlength{\itemindent}{0pt}
  \setlength{\leftmargin}{0pt}
  \setlength{\parsep}{0pt}
  \ifodd #1
   \setlength{\leftmargin}{\cslhangindent}
   \setlength{\itemindent}{-1\cslhangindent}
  \fi
  \setlength{\itemsep}{#2\baselineskip}}}
 {\end{list}}
\usepackage{calc}

\newcommand{\CSLLeftMargin}[1]{\parbox[t]{\csllabelwidth}{\strut#1\strut}}
\newcommand{\CSLRightInline}[1]{\parbox[t]{\linewidth - \csllabelwidth}{\strut#1\strut}}

\providecommand{\tightlist}{%
  \setlength{\itemsep}{0pt}\setlength{\parskip}{0pt}}
\usepackage{booktabs}
\usepackage{longtable}
\usepackage{array}
\usepackage{multirow}
\usepackage{wrapfig}
\usepackage{float}
\usepackage{colortbl}
\usepackage{pdflscape}
\usepackage{tabu}
\usepackage{threeparttable}
\usepackage{threeparttablex}
\usepackage[normalem]{ulem}
\usepackage{makecell}
\usepackage{xcolor}
\usepackage{bookmark}
\IfFileExists{xurl.sty}{\usepackage{xurl}}{} 
\makeatletter
\@ifundefined{xmpquote}{\newcommand{\xmpquote}[1]{#1}}{}
\makeatother
\hypersetup{
  pdftitle={Which Regularized Propensity-Score and Doubly Robust Methods Are Best Calibrated When Exposures or Outcomes Are Rare? A Plasmode Study of Proxy-Based Confounding Adjustment},
  pdfauthor={M. Ehsan Karim; Wanqing Hu},
  pdfkeywords={\xmpquote{causal inference; proxy-based confounding
adjustment; high-dimensional propensity score; event rarity; LASSO;
targeted maximum likelihood estimation; plasmode simulation}},
  hidelinks,
  pdfcreator={LaTeX via pandoc}}

\title{Which Regularized Propensity-Score and Doubly Robust Methods Are
Best Calibrated When Exposures or Outcomes Are Rare? A Plasmode Study of
Proxy-Based Confounding Adjustment}
\author{M. Ehsan Karim\footnote{School of Population and Public Health,
  University of British Columbia, Vancouver, Canada. Corresponding
  author:
  \href{mailto:ehsan.karim@ubc.ca}{\nolinkurl{ehsan.karim@ubc.ca}}} \and Wanqing
Hu\footnote{School of Population and Public Health, University of
  British Columbia, Vancouver, Canada}}
\date{July 2026}

\begin{document}
\maketitle
\begin{abstract}
\textbf{Purpose.} Confounding adjustment in health-care database studies
screens large proxy libraries where events per variable are low,
straining standard propensity score (PS) methods, and many regularized
variable-selection strategies exist (outcome-adaptive LASSO {[}OAL{]},
group LASSO/GLiDeR, highly adaptive LASSO {[}HAL{]}). Yet few
comparisons have varied exposure prevalence within such a selection
menu, and none pairs it with doubly robust estimation, compute
accounting, and a null-RD truth anchor.

\textbf{Methods.} We conducted a plasmode simulation anchored on
National Health and Nutrition Examination Survey data (2013--2018; 25
investigator-specified covariates, 142 prescription-derived proxies),
comparing ten pipelines combining these strategies with inverse
probability of treatment weighting (IPTW) and targeted maximum
likelihood estimation (TMLE). Three scenarios were evaluated under a
known null (true risk difference, RD = 0): frequent, rare-exposure, and
rare-outcome. We report bias, standard error (SE), relative error, 95\%
coverage, and runtime.

\textbf{Results.} HAL (G-Computation) had near-zero bias but highly
concentrated estimates, giving near-unity coverage and large relative
error (106--186\%). OAL (IPTW), GLiDeR, and HAL (TMLE) were best
calibrated, whereas the regularized-LASSO TMLE pipelines under-covered
modestly (91--93\%) in the rare scenarios. Under rare exposure,
LASSO-IPTW had the largest bias and inflated SE and over-covered
(conservatively), problems that TMLE removed. On real data, methods
agreed (RD approximately 0.07--0.085). Runtimes spanned \textless1 s to
\textgreater16 h.

\textbf{Conclusions.} Under a null benchmark, pairing outcome-aware
selection (OAL, GLiDeR) or doubly robust estimation (TMLE) with
regularized models best balanced bias, calibration, and robustness to
rarity. The rare-exposure arm exposed the largest gaps; method choice
should weigh the prioritized metric against compute.
\end{abstract}

\emph{Running title:} Regularized confounding-adjustment comparison

\section*{Key Points}\label{key-points}
\addcontentsline{toc}{section}{Key Points}

\begin{itemize}
\tightlist
\item
  Under one truth-anchored plasmode design we varied exposure
  prevalence---rarely examined within such a selection-and-estimator
  menu---while comparing ten proxy-based confounding-adjustment
  pipelines (traditional LASSO, collaborative-controlled LASSO,
  outcome-adaptive LASSO {[}OAL{]}, group LASSO/GLiDeR, and highly
  adaptive LASSO {[}HAL{]}, each paired with IPTW and/or TMLE) under a
  known null effect.
\item
  Outcome-aware selection (OAL, GLiDeR) and doubly robust estimation
  (HAL-TMLE) were best calibrated (OAL coverage 94.7--95.8\% across
  scenarios; GLiDeR and HAL-TMLE close behind), whereas HAL
  (G-Computation) had near-zero bias but over-concentrated estimates
  with over-covering intervals (relative error 106--186\%).
\item
  Under rare exposure, LASSO-IPTW had the largest bias and over-covered
  conservatively; replacing IPTW with TMLE removed the bias and
  instability, at the cost of mild undercoverage (91--93\%).
\item
  The best-calibrated methods were also the slowest (under 1 second to
  over 16 hours per analysis, single core), so method choice must weigh
  statistical performance against compute.
\end{itemize}

\section*{Plain Language Summary}\label{plain-language-summary}
\addcontentsline{toc}{section}{Plain Language Summary}

When researchers use health records to ask whether one factor causes a
health outcome, they must account for many background characteristics
that could distort the answer. Modern datasets can contain hundreds of
such characteristics, more than older statistical tools can handle
reliably, so newer ``variable-selection'' methods are needed to pick the
ones that matter. We tested ten such methods on realistic simulated data
built from a large US national health survey---keeping people's real
background characteristics but simulating the studied factor and the
outcome---where we knew in advance that the studied factor had no true
effect, so any apparent effect was an error we could measure. We also
applied all ten methods to real data on the link between obesity and
type 2 diabetes. Methods that focused on characteristics linked to the
outcome, or that combined two modelling steps for extra robustness, gave
the most trustworthy results in the settings we tested, with error
margins that were neither too wide nor too narrow. One popular simpler
method performed poorly when the studied factor was uncommon, but a more
advanced version fixed this. The most reliable methods were also the
slowest, taking hours rather than seconds, so researchers must balance
accuracy against computing time when choosing a method.

\section{Introduction}\label{introduction}

\subsection{Confounding adjustment with large proxy
sets}\label{confounding-adjustment-with-large-proxy-sets}

Causal effect estimation from observational data is threatened by
confounding, and in pharmacoepidemiology and health-care database
studies the adjustment set is often very large. Beyond
investigator-specified covariates, routinely collected \emph{proxy}
variables---administrative records such as pharmacy dispensings or
diagnostic codes---can capture otherwise unmeasured confounders and
reduce residual bias\textsuperscript{1}. When the adjustment set holds
many correlated proxies while the number of outcome or exposure events
is modest---so events per variable are low (here as few as roughly 2.5
under rare outcomes)---standard maximum-likelihood propensity score (PS)
models become unstable, and regularization or variable selection becomes
essential\textsuperscript{2,3}. This regime is the same whether proxies
come from claims dispensings, electronic-health-record codes, or a
national survey's prescription records; we use the publicly available
National Health and Nutrition Examination Survey
(NHANES)\textsuperscript{4} as a reproducible anchor for a plasmode
simulation, which keeps real covariate data while imposing a known
effect.

A large methodological literature has responded. The high-dimensional
propensity score (hdPS) and its machine-learning
extensions\textsuperscript{2,3} adapt covariate prioritization to large
proxy libraries, a fast-growing toolkit recently reviewed for
pharmacoepidemiology\textsuperscript{5}. Within the LASSO family,
variants differ in \emph{what} the penalty is designed to
select---collaborative-controlled LASSO\textsuperscript{6,7}, the
outcome-adaptive LASSO (OAL)\textsuperscript{8}, group LASSO with doubly
robust estimation (GLiDeR)\textsuperscript{9}, and the highly adaptive
LASSO (HAL)\textsuperscript{10,11}---and are typically combined with a
downstream estimator, most commonly inverse probability of treatment
weighting (IPTW) or the doubly robust targeted maximum likelihood
estimator (TMLE).

\subsection{Aim and contributions}\label{aim-and-contributions}

Despite this breadth, each method was largely validated in
isolation---under balanced, low-dimensional, or otherwise favourable
conditions and on heterogeneous estimands---with no common design. Where
selection has been crossed with a downstream estimator, the comparison
stayed narrow: OAL versus its generalization, crossed with IPTW, its
augmented form (AIPW), and TMLE but only under extremal-propensity
overlap\textsuperscript{12}, or HAL-family pipelines on low-dimensional
synthetic data\textsuperscript{13,14}. Crucially, exposure prevalence
has rarely been varied, even though reviews flag proxy adjustment as
unstable in studies with rare events\textsuperscript{5}. We provide the
missing common-design, truth-anchored comparison that assembles these
axes---selection breadth, weighting versus doubly robust estimation,
exposure- and outcome-rarity, and compute---which no prior study holds
at once. Our contributions are to:

\begin{enumerate}
\def\labelenumi{\arabic{enumi}.}
\tightlist
\item
  unify five regularized selection families---traditional LASSO,
  collaborative-controlled LASSO, OAL, group LASSO (GLiDeR), and
  HAL---each paired with IPTW and/or doubly robust TMLE under one
  truth-anchored design (true risk difference, RD = 0); to our knowledge
  no prior single-design study has assembled this breadth;
\item
  introduce a rare-\emph{exposure} arm---varying treatment prevalence,
  which prior proxy-adjustment comparisons have rarely
  stressed---alongside a rare-outcome arm that confirms, under a broader
  method set, the instability documented by Wyss et
  al.\textsuperscript{7};
\item
  assess whether replacing IPTW with a doubly robust TMLE step improves
  performance (a within-selection swap tested directly for the
  traditional LASSO pipelines, and as a weighting-versus-doubly-robust
  contrast across the wider menu); and
\item
  quantify each pipeline's computational cost---from under a second to
  over sixteen hours per single-core analysis, unreported by prior
  comparisons---to inform practical method choice.
\end{enumerate}

To complement the simulation, we also apply all methods to real-world
NHANES data covering 2013--2018, as an illustrative analysis of the
obesity--type 2 diabetes association.

\section{Methods}\label{methods}

\subsection{Notation and estimand}\label{notation-and-estimand}

Let \((A, Y, L, Z)\) denote, for \(n\) independent units, a binary
treatment \(A\) (obesity), a binary outcome \(Y\) (diabetes),
investigator-specified baseline covariates \(L\), and proxy variables
\(Z\). Writing \(Y(a)\) for the potential outcome under treatment
\(a \in \{0,1\}\), we target the marginal \textbf{risk difference}
\[ \mathrm{RD} = \mathbb{E}[Y(1)] - \mathbb{E}[Y(0)]. \] Identification
of the RD from the observed data relies on the standard assumptions of
(i) conditional exchangeability, \(\{Y(0), Y(1)\} \perp A \mid (L, Z)\);
(ii) positivity, \(0 < \mathbb{P}(A = 1 \mid L, Z) < 1\); (iii)
consistency; and (iv) no interference (the stable-unit-treatment-value
assumption, SUTVA). Proxy adjustment aims to make exchangeability more
plausible by capturing otherwise unmeasured confounders, without
guaranteeing it. All methods estimate the propensity score
\(\pi(L, Z) = \mathbb{P}(A = 1 \mid L, Z)\) and/or the outcome
regression \(Q(A, L, Z) = \mathbb{E}[Y \mid A, L, Z]\) using regularized
models.

\subsection{Generic estimators}\label{generic-estimators}

We pair each selection strategy with one or both estimators.
\textbf{IPTW} reweights subjects by the inverse of their estimated
treatment probability---consistent for the RD under exchangeability and
positivity, but sensitive to extreme weights near positivity
violations\textsuperscript{15--17}. \textbf{TMLE} is a doubly robust,
semiparametric-efficient plug-in estimator that updates an initial
outcome regression with a PS-driven targeting step and quantifies
uncertainty through the efficient influence curve (the estimator's
first-order sampling behavior, from which its standard error is
derived), remaining consistent if either nuisance model is
correct\textsuperscript{18--20}. Thus ``LASSO-IPTW'' and ``LASSO-TMLE''
denote identical nuisance models with different downstream estimators.

\subsection{Candidate methods}\label{candidate-methods}

We compare ten pipelines built from five regularized \emph{selection}
strategies, each paired with one or two downstream \emph{estimators}
(Table 1). A traditional LASSO baseline selects proxies by a
cross-validated outcome LASSO (under deviance and MSE loss) and fits an
unpenalized propensity score (PS) estimated by IPTW and by
TMLE\textsuperscript{2,3}. The collaborative-controlled family fits
LASSO PS models and applies TMLE either with a collaborative
undersmoothing search (C-TMLE) or at the cross-validation choice
(Standard TMLE)\textsuperscript{6,7}. The outcome-adaptive LASSO (OAL)
penalizes covariates by their outcome association, preferentially
retaining confounders\textsuperscript{8}. Group LASSO with doubly robust
estimation (GLiDeR) links selection across the outcome and treatment
models\textsuperscript{9}. The highly adaptive LASSO (HAL) applies
\(L_1\) regularization to a rich basis expansion of the outcome and
enters as G-computation and as TMLE\textsuperscript{10,11}. Propensity
scores in the LASSO-based and TMLE pipelines are truncated at
\(5/(\sqrt{n}\,\log n)\)\textsuperscript{20,21}; OAL and GLiDeR weights
follow their published untruncated form (Table S1), so their stability
reflects selection rather than truncation. Bootstrap intervals use 200
percentile resamples. Full implementation and tuning are in Web Appendix
A and Table S1, and the OAL tuning-parameter range is justified in Web
Appendix B.

\begin{table}[H]
\centering
\caption{\label{tab:tab1-methods}The ten estimation pipelines: five regularized selection strategies paired with downstream estimators, with interval construction, the key methods reference(s), and the R packages used (the LASSO baseline is run under deviance and MSE loss, giving two pipelines each for IPTW and TMLE). Bootstrap inference uses 200 percentile resamples. References are given as author-year for readability and correspond to the numbered entries in the reference list; tuning parameters are in Table S1 and detailed implementation in Web Appendix A.}
\centering
\fontsize{8}{10}\selectfont
\begin{tabular}[t]{>{\raggedright\arraybackslash}p{2.0cm}>{\raggedright\arraybackslash}p{3.1cm}>{\raggedright\arraybackslash}p{1.5cm}>{\raggedright\arraybackslash}p{1.7cm}>{\raggedright\arraybackslash}p{2.6cm}>{\raggedright\arraybackslash}p{2.3cm}}
\toprule
\cellcolor[HTML]{E8ECF0}{Pipeline} & \cellcolor[HTML]{E8ECF0}{Selection (penalty target)} & \cellcolor[HTML]{E8ECF0}{Estimator} & \cellcolor[HTML]{E8ECF0}{Inference} & \cellcolor[HTML]{E8ECF0}{Key reference(s)} & \cellcolor[HTML]{E8ECF0}{R package}\\
\midrule
LASSO (IPTW) & CV outcome-LASSO proxies (deviance, MSE) & IPTW & Sandwich Wald & Karim 2018, 2025 & glmnet, sandwich\\
LASSO (TMLE) & CV outcome-LASSO proxies (deviance, MSE) & TMLE & Influence curve & Karim 2018, 2025 & glmnet, tmle\\
C-TMLE (Collab.) & Undersmoothed LASSO PS & C-TMLE & Influence curve & Ju 2019; Wyss 2024 & ctmle\\
Standard TMLE & CV-LASSO PS & TMLE & Influence curve & Ju 2019; Wyss 2024 & tmle, glmnet\\
OAL (IPTW) & Outcome-adaptive LASSO PS & IPTW & Bootstrap & Shortreed 2017 & OAL source, glmnet\\
\addlinespace
GLiDeR & Adaptive group LASSO (joint) & Doubly robust & Bootstrap & Koch 2018 & GLiDeR source\\
HAL (G-Comp.) & HAL basis (outcome) & G-computation & Bootstrap & Benkeser 2016; Butzin-Dozier 2024 & hal9001\\
HAL (TMLE) & HAL basis (outcome + PS) & TMLE & Influence curve & Benkeser 2016; Butzin-Dozier 2024 & hal9001, tmle\\
\bottomrule
\end{tabular}
\end{table}

\subsection{Motivating empirical
example}\label{motivating-empirical-example}

Excess body fat drives type 2 diabetes through disrupted insulin
signaling and glucose metabolism\textsuperscript{22}. Our empirical
analysis uses the processed NHANES dataset of Karim and
Lei\textsuperscript{1}: 7,585 individuals with 14 categorical baseline
characteristics and 11 continuous physiological measures. Proxy
variables derived from the NHANES Prescription Medications component may
capture aspects of comorbidity burden and health-care use not otherwise
recorded in the survey, potentially reducing unmeasured confounding. We
analyzed this processed complete-case cohort without NHANES survey
weights, strata, or primary sampling units; the resulting risk
differences are therefore unweighted, cross-sectional associations that
demonstrate pipeline feasibility on genuine prescription-derived proxy
data rather than nationally representative or causal effects.

\subsection{Plasmode simulation}\label{plasmode-simulation}

We evaluated the methods using a plasmode
simulation\textsuperscript{23}, which preserves the empirical covariate
dependence of real data while imposing a known truth---now a standard
approach for evaluating proxy-based confounder-adjustment
methods\textsuperscript{1,7,24}---and we report the design following the
ADEMP framework (Aims, Data-generating mechanisms, Estimands, Methods,
Performance measures)\textsuperscript{25}. The generating procedure fits
outcome and exposure models on the anchor data and then, for each
resampled individual, draws a new exposure and outcome from these models
under an imposed effect, leaving the real covariate values and their
correlations intact; the software implementation and exact generating
call are in Web Appendix F.

The plasmode was anchored on the processed NHANES 2013--2018 dataset of
Karim and Lei\textsuperscript{1} (N = 7,585; 25 investigator-specified
covariates and 142 prescription-derived proxies). The procedure fit
logistic outcome (\(\mathrm{outcome} \sim \mathrm{exposure} + L\)) and
exposure (\(\mathrm{exposure} \sim L\)) base models, where \(L\)
collects the 14 categorical covariates, six laboratory transforms
derived from the 11 continuous measures, and a proxy-burden count
summarizing the 94 of the 142 proxies that were outcome-associated. Each
replicate resamples individuals with replacement and draws new exposure
and outcome values from these models; the empirical covariate and proxy
\emph{values}---and hence their correlation structure---carry through
unchanged, while the generative \emph{coefficients} were set
deliberately. We configured the procedure to generate exposure from the
fitted propensity model (\(\mathrm{exposure} \sim L\)) rather than
carrying the observed treatment through unchanged---the
generate-treatment design of Shaw et al.\textsuperscript{26} that avoids
the positivity violation which can otherwise spuriously penalize
propensity-score and weighting estimators in plasmode evaluations; the
exact generating call is reproduced in Web Appendix F. Each replicate's
simulated exposure and outcome were then re-joined to the full set of
investigator covariates and all 142 proxies, the candidate adjustment
set supplied to every method.

The exposure--outcome effect was fixed at an odds ratio of one, so the
true marginal RD is \textbf{zero}. The outcome-model coefficient on the
proxy-burden count was scaled fivefold, making the cumulative
proxy-burden count a strong confounder by design while the exposure
effect remained null; this engineered confounding is what the methods
must adjust away and underlies the overlap behavior in Table S4. Fixing
the effect at the null with a strong, known confounder is a deliberate
calibration design: any nonzero estimate is pure residual-confounding
bias, which isolates adjustment quality from effect estimation. Bias is
therefore \(\overline{\widehat{\mathrm{RD}}}\), and 95\% coverage
targets the nominal level under this engineered null, which differs from
the positive obesity--diabetes association in the real data (Figure 1).
For each scenario we generated 1,000 replicates of \(n = 4{,}500\) using
a fixed seed.

Three scenarios contrast event rarity via the target exposure and
outcome prevalences set in the generating procedure (Table 2): frequent
exposure and outcome; rare exposure; and rare outcome. Rarity was
induced by shifting the exposure- and outcome-model intercepts to hit
each target while holding the covariate-driven coefficients fixed---not
by subsampling---so realized prevalences vary slightly (Table 2).

\begin{table}[H]
\centering
\caption{\label{tab:tab1-scenarios}Plasmode simulation scenarios: realized exposure and outcome prevalences. All scenarios use n = 4,500 per replicate, 1,000 replicates, a true risk difference of 0, and 167 candidate predictors (25 investigator-specified covariates and 142 prescription-derived proxy variables); the effective number entering each model varies after frequency filtering and dummy coding.}
\centering
\begin{tabular}[t]{lll}
\toprule
\cellcolor[HTML]{E8ECF0}{Scenario} & \cellcolor[HTML]{E8ECF0}{Exposure prev.} & \cellcolor[HTML]{E8ECF0}{Outcome prev.}\\
\midrule
Frequent & 28.5\% & 29.2\%\\
Rare exposure & 9.4\% & 28.2\%\\
Rare outcome & 28.5\% & 9.3\%\\
\bottomrule
\end{tabular}
\end{table}

\subsection{Performance measures}\label{performance-measures}

With true value \(\theta = 0\), we summarized performance across
replicates with the \texttt{rsimsum} package\textsuperscript{27}
following Morris et al.\textsuperscript{25}: \textbf{bias},
\textbf{empirical SE}, \textbf{model-based SE}, \textbf{relative error}
of the model SE (positive values indicate overstated variability), and
\textbf{95\% coverage}. Formal definitions of each measure are given in
Web Appendix C. Coverage uses the Wald interval
\(\widehat{\mathrm{RD}}_k \pm 1.96\,\widehat{\mathrm{SE}}_k\) from each
method's reported SE (the bootstrap SD for bootstrap-based methods),
applied uniformly so calibration reflects the reported SE; the
percentile bootstrap intervals are used for the methods' own point
estimates and the real-data analysis. We also summarized overlap and
positivity for the LASSO-fitted PS per replicate, averaged within each
scenario (Table S4).

All comparative metrics are computed on the 966 replicates common to
every method (of 1,000 per scenario), so differences reflect the methods
rather than different samples. The occasional bootstrap non-convergence,
the robustness of this common-replicate choice, and the Monte Carlo
error (about 0.7 percentage points for a coverage near 95\%) are
detailed in Web Appendix G.

\subsection{Software}\label{software}

All analyses used R. The selection and estimation package for each
pipeline is listed in Table 1; the plasmode replicates were generated
with the \texttt{Plasmode} package (Web Appendix F), performance metrics
with \texttt{rsimsum}\textsuperscript{27}, and replicates were processed
in parallel with \texttt{future}. All method-implementation and
simulation code is publicly available (see the Transparency and
Reproducibility statement).

\section{Results}\label{results}

\subsection{Motivating example}\label{motivating-example}

\begin{figure}

{\centering \includegraphics[width=0.65\linewidth]{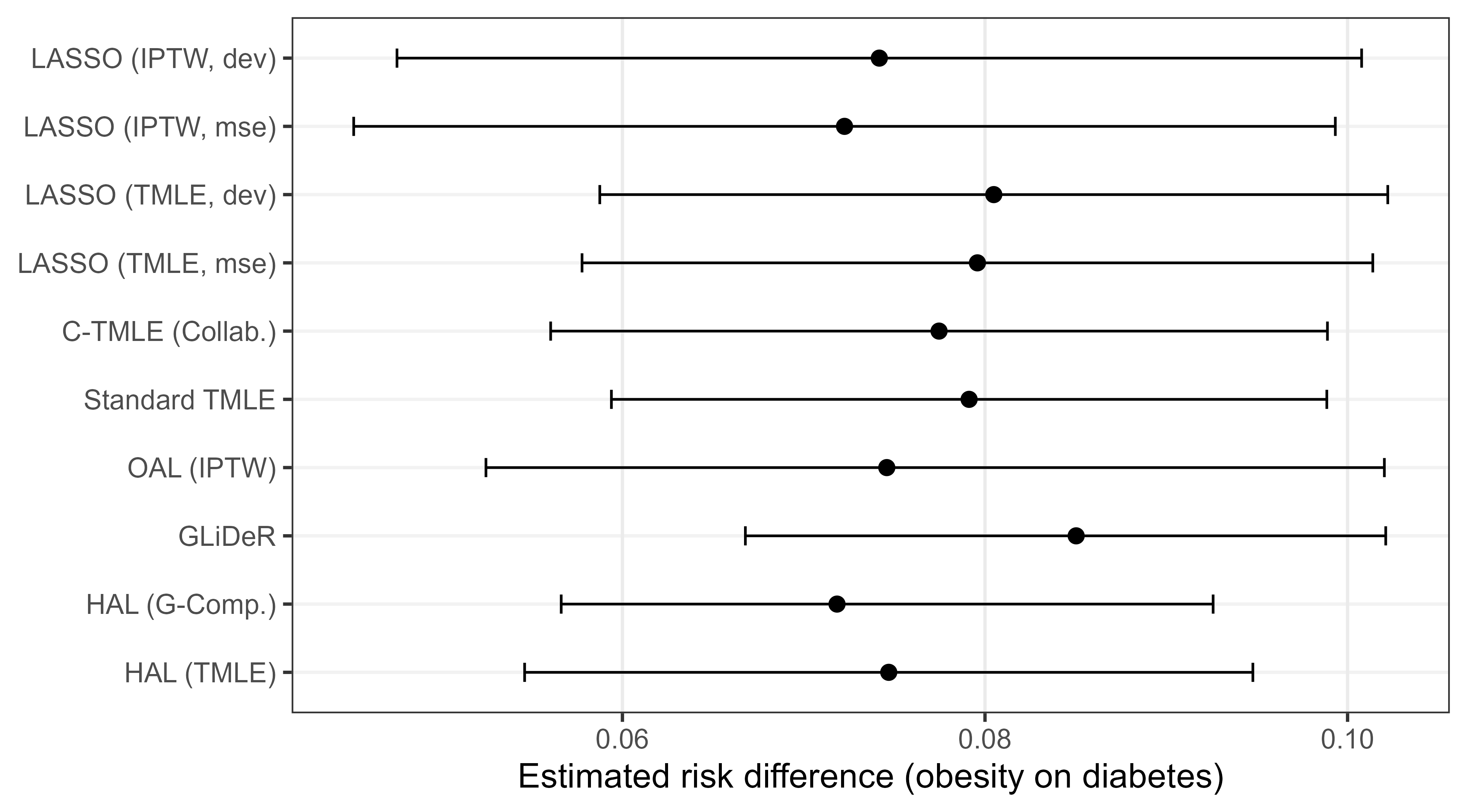} 

}

\caption{Estimated risk difference for the association between obesity and diabetes in the NHANES data, by method. Pipeline labels combine the selection strategy with its downstream estimator and, for the traditional LASSO baseline, the cross-validation loss used to select proxies: `IPTW' vs.\ `TMLE' names the estimator and `dev' vs.\ `mse' names deviance vs.\ mean-squared-error loss---so, e.g., LASSO (IPTW, dev) is LASSO proxy selection under deviance loss estimated by IPTW. Intervals are 95\% bootstrap percentile intervals for OAL (IPTW), GLiDeR, and HAL (G-Computation); efficient-influence-curve intervals for the TMLE methods; and a model-based Wald interval for LASSO-IPTW; interval widths are therefore not strictly comparable across methods.}\label{fig:fig1-motivating}
\end{figure}

Most methods produced RD estimates ranging from approximately 0.07 to
0.085, with GLiDeR yielding the highest estimate and HAL (G-Computation)
the lowest (Figure 1). Overall, the point estimates were relatively
consistent across methods, and their confidence intervals showed
substantial overlap, indicating that---on these real data---the choice
among methods had limited impact on the qualitative conclusion of a
positive obesity--diabetes association. This panel is a feasibility and
consistency check---the pipelines run end-to-end on genuine
prescription-derived proxy data and agree---rather than a performance
benchmark, since the truth is unknown here and the interval widths are
not strictly comparable.

\subsection{Plasmode simulation
results}\label{plasmode-simulation-results}

Table 3 reports the headline performance---bias, 95\% coverage, and the
relative error of the model-based SE---for every method in each
scenario, organized so that a method can be read across the three rarity
settings; Figure 2 displays the coverage with Monte Carlo intervals.
Bias, empirical SE, and relative error are shown graphically in
Supporting Information (Figures S1--S3), and the full table including
empirical and model-based SEs in Table S2. Table 3 and the figures
derive from a single \texttt{rsimsum} summary of the saved
replicate-level results.

\begin{table}[H]
\centering
\caption{\label{tab:tab2-results}Simulation performance on the 966 replicates common to all methods (true RD = 0), arranged so each method can be read across the three rarity settings. Bias is $\times 10^{-3}$; Cov.\ is 95\% Wald coverage (\%, target 95); RE is the relative error of the model-based SE (\%); per scenario, all coverages within Monte Carlo error of nominal ($|\Delta|\le 1.5$ pp) are bold. Empirical and model-based SEs and Monte Carlo errors are in Supporting Information (Table S2).}
\centering
\fontsize{9}{11}\selectfont
\begin{tabular}[t]{lrrrrrrrrr}
\toprule
\multicolumn{1}{c}{ } & \multicolumn{3}{c}{Frequent} & \multicolumn{3}{c}{Rare exposure} & \multicolumn{3}{c}{Rare outcome} \\
\cellcolor[HTML]{E8ECF0}{Method} & \cellcolor[HTML]{E8ECF0}{Bias} & \cellcolor[HTML]{E8ECF0}{Cov.} & \cellcolor[HTML]{E8ECF0}{RE} & \cellcolor[HTML]{E8ECF0}{Bias} & \cellcolor[HTML]{E8ECF0}{Cov.} & \cellcolor[HTML]{E8ECF0}{RE} & \cellcolor[HTML]{E8ECF0}{Bias} & \cellcolor[HTML]{E8ECF0}{Cov.} & \cellcolor[HTML]{E8ECF0}{RE}\\
\midrule
LASSO (IPTW, dev) & 0.17 & 99.3 & 38 & -5.20 & 97.7 & 22 & 0.33 & 98.7 & 27\\
LASSO (IPTW, mse) & 0.18 & 99.3 & 38 & -5.17 & 97.8 & 22 & 0.32 & 98.7 & 27\\
LASSO (TMLE, dev) & -0.20 & \textbf{94.6} & -4 & -0.11 & 91.8 & -8 & 0.10 & 91.6 & -12\\
LASSO (TMLE, mse) & -0.19 & \textbf{94.7} & -4 & -0.10 & 92.0 & -8 & 0.12 & 91.7 & -12\\
C-TMLE (Collab.) & -0.15 & \textbf{94.2} & -3 & -0.98 & 91.3 & -14 & -0.03 & 92.1 & -9\\
\addlinespace
Standard TMLE & -0.13 & \textbf{93.9} & -4 & -0.04 & 92.4 & -7 & 0.02 & 92.5 & -8\\
OAL (IPTW) & -0.30 & \textbf{95.8} & 2 & -1.42 & \textbf{94.9} & 1 & 0.09 & \textbf{94.7} & 2\\
GLiDeR & 0.49 & \textbf{94.6} & 0 & 0.87 & \textbf{93.9} & -4 & 0.34 & \textbf{93.7} & -2\\
HAL (G-Comp.) & -0.08 & 99.9 & 106 & -0.09 & 99.9 & 186 & -0.00 & 99.9 & 130\\
HAL (TMLE) & -0.13 & \textbf{94.9} & -1 & -0.08 & \textbf{94.0} & -4 & 0.02 & \textbf{94.6} & -1\\
\bottomrule
\end{tabular}
\end{table}

\begin{figure}

{\centering \includegraphics[width=0.92\linewidth]{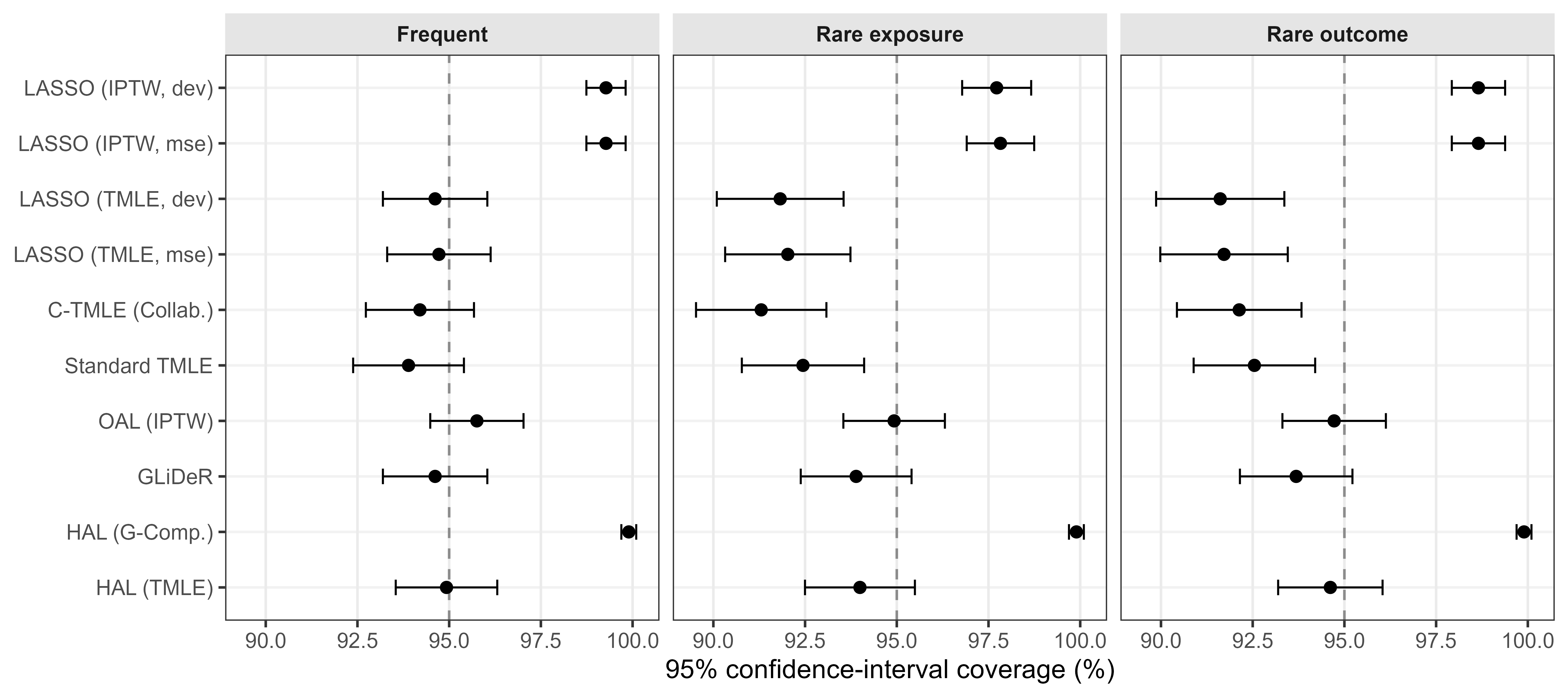} 

}

\caption{95\% confidence-interval coverage by method under the frequent, rare-exposure, and rare-outcome scenarios; the dashed line marks the nominal 95\% level. Bias, empirical standard error, and relative error appear in Supporting Information (Figures S1--S3).}\label{fig:fig2-cover}
\end{figure}

\subsubsection{Frequent exposure and
outcome}\label{frequent-exposure-and-outcome}

Point estimates were close to the truth for all methods (Table 3; Figure
S1). Calibration differed more: HAL (G-Computation) substantially
over-covered (99.9\%), as did both LASSO-IPTW pipelines (approximately
99.3\%), while most others were near nominal (94.2--95.8\%)---HAL
(TMLE), the LASSO-TMLE pipelines, and GLiDeR closest, and OAL (IPTW)
slightly conservative at 95.8\% (Figure 2). HAL (G-Computation) had by
far the smallest empirical SE and correspondingly the largest relative
error (106\%), because its bootstrap SE greatly exceeds its tiny
empirical SD; the TMLE-based methods, GLiDeR, and OAL had small relative
errors (Figure S3).

\subsubsection{Rare exposure and frequent
outcome}\label{rare-exposure-and-frequent-outcome}

This scenario was the most discriminating. Both LASSO-IPTW pipelines
showed the largest bias (\(-5.2 \times 10^{-3}\)) and empirical SE
(0.022), reflecting weighting instability under limited overlap even
after truncation, and over-covered (approximately 97.7--97.8\%).
Replacing IPTW with TMLE essentially removed this: the LASSO-TMLE
pipelines had negligible bias (approximately \(-1 \times 10^{-4}\)) and
lower empirical SE (approximately 0.018), though coverage fell slightly
below nominal (approximately 91.8--92.0\%). OAL (IPTW) was best
calibrated (coverage 94.9\%, relative error 1.2\%), with HAL (TMLE)
(94.0\%) and GLiDeR (93.9\%) close; the remaining TMLE methods
under-covered modestly (approximately 91.3--92.4\%). HAL (G-Computation)
again combined near-zero bias and the smallest empirical SE (0.0038)
with near-unity coverage (99.9\%) and the largest relative error
(185.9\%). Overlap diagnostics (Table S4) confirm the positivity stress:
the largest weight rose to about 56 (versus 16 when exposure was
frequent) and about 2\% of scores were truncated, more so for the
collaborative than the standard model (Web Appendix D).

\subsubsection{Frequent exposure and rare
outcome}\label{frequent-exposure-and-rare-outcome}

Patterns echoed the rare-exposure scenario. HAL (G-Computation) had the
smallest bias (\(-3.8 \times 10^{-7}\)) and HAL (TMLE) the next
(\(1.6 \times 10^{-5}\)). Coverage was again closest to nominal for OAL
(IPTW) (94.7\%) and HAL (TMLE) (94.6\%), with the other TMLE methods
lower (approximately 91.6--92.5\%) and LASSO-IPTW and HAL
(G-Computation) markedly over-covering (approximately 98.7\% and
99.9\%). HAL (G-Computation) again had the smallest empirical SE
(0.0017) and largest relative error (129.8\%), while GLiDeR, HAL (TMLE),
and OAL had relative errors \(\le 2.1\%\) in magnitude (Figure S3).

\subsection{Computing time}\label{computing-time}

\begin{table}[H]
\centering
\caption{\label{tab:tab4-runtime}Single-core runtime per method for one analysis of the NHANES data, ordered fastest to slowest. Each figure is the per-analysis wall-clock time on a single core: the R = 200 non-parametric bootstrap was run serially, so these equal each method's CPU cost. They are NOT aggregates over the 1,000-replicate simulation (which instead parallelized replicates across the compute node). The slow methods are dominated by the bootstrap; the portable message is the order-of-magnitude penalty of bootstrap over influence-curve inference rather than the exact ranking.}
\centering
\begin{tabular}[t]{ll}
\toprule
\cellcolor[HTML]{E8ECF0}{Method} & \cellcolor[HTML]{E8ECF0}{Runtime}\\
\midrule
LASSO (IPTW, mse) & 0.07 s\\
LASSO (IPTW, dev) & 0.31 s\\
Standard TMLE & 1.46 s\\
LASSO (TMLE, mse) & 4.90 s\\
LASSO (TMLE, dev) & 5.38 s\\
\addlinespace
C-TMLE (Collab.) & 12.22 s\\
HAL (TMLE) & 6.8 min\\
OAL (IPTW) & 2.2 h\\
GLiDeR & 7.3 h\\
HAL (G-Comp.) & 16.3 h\\
\bottomrule
\end{tabular}
\end{table}

Per-analysis runtime spanned nearly six orders of magnitude (Table 4).
The weighting and standard TMLE pipelines were fastest (under a second
to a few seconds) and HAL (TMLE) took minutes, whereas the
bootstrap-based methods were far more expensive---OAL about 2 hours,
GLiDeR about 7 hours, HAL (G-Computation) over 16 hours on a single
core---with the R = 200 serially executed non-parametric bootstrap the
dominant bottleneck.

\section{Discussion}\label{discussion}

\subsection{Key findings}\label{key-findings}

Our four aims yield four findings. \textbf{First}, across the common
design, methods with more targeted selection (OAL, GLiDeR) or doubly
robust estimation (TMLE) were better calibrated with smaller residual
bias than the traditional LASSO--IPTW baseline, though no single method
dominated every metric. \textbf{Second}, event rarity was the decisive
stressor: performance was similar when events were frequent but diverged
under rare exposure, where the weighting-based pipelines destabilized.
\textbf{Third}, replacing IPTW with a doubly robust TMLE step
consistently improved bias and precision for a given selection strategy,
at a small cost in coverage under rarity. \textbf{Fourth}, per-analysis
computational cost spanned nearly six orders of magnitude---from under a
second to over sixteen hours (HAL G-Computation, single core)---and was
largely inversely related to calibration, so the most appropriate choice
depends on which property---bias, calibration, precision, or
compute---is prioritized (Table 5).

\subsection{Comparison with prior
work}\label{comparison-with-prior-work}

All ten pipelines extend the high-dimensional propensity score (hdPS)
paradigm\textsuperscript{28}, replacing its univariate proxy ranking
with joint, outcome-aware regularized selection\textsuperscript{5}.
Prior benchmarks in this tradition pitted hdPS against a \emph{single}
alternative---regularized outcome regression\textsuperscript{24}---or
against a handful of machine-learning selection methods and hdPS
hybrids\textsuperscript{2}; we instead cross a five-family menu with
IPTW versus doubly robust estimation. Each origin study, moreover,
validated its method against a \emph{different, self-favouring}
comparator set: no prior work places OAL, GLiDeR, C-TMLE, and HAL in one
design, nor reports comparative runtime (Supporting Information Table
S3).

Our closest antecedents temper, rather than negate, the rarity
contribution. Rassen et al.~examined hdPS with few exposed and few
events\textsuperscript{29}. Our closest published sibling, Franklin et
al.\textsuperscript{30}, shares our plasmode scaffolding and known null
but varied propensity-score \emph{utilization} (matching,
stratification, weighting) for the risk ratio---not the regularized
\emph{selection} families we compare---and excluded doubly robust
estimators a priori, reasoning that outcome modeling fails with few
events. We instead evaluate doubly robust TMLE empirically under rarity,
and scope our exposure-prevalence claim to this
selection-menu-versus-estimator design anchored to a null risk
difference. Jones et al.~found the efficient estimator dominates the
OAL-versus-generalization selection choice under near-positivity driven
by covariate correlation, crossing an overlap-weighting axis we hold
fixed\textsuperscript{12}; Wyss et al.~already benchmark our LASSO, OAL,
and collaborative-controlled arms with IPTW, AIPW, and TMLE but fix
treatment prevalence at 40\% on a rare outcome\textsuperscript{7}---the
rare-outcome lessons are theirs, not ours. Relative to both we add
GLiDeR and HAL, a rare-\emph{exposure} arm, and runtime. OAL is moreover
known to lose stability under correlated covariates, motivating the
generalized OAL\textsuperscript{31}---untested for our collinear
proxies---so OAL's strong calibration here likely reflects dominant
baseline confounders. Finally, the outcome-adaptive HAL line fuses these
ingredients into one estimator\textsuperscript{13,14}; our menu
benchmarks them as separate arms at a scale where the fused estimator is
infeasible.

\subsection{Method-specific
observations}\label{method-specific-observations}

Under rare exposure the LASSO--IPTW methods showed large bias and
inflated empirical SE even with truncation, yet over-covered: the
conservative sandwich SE (which treats the weights as fixed) inflates
interval width enough to over-cover despite the off-centering, unlike
OAL's well-calibrated bootstrap IPTW. Replacing IPTW with TMLE markedly
improved bias and variability, though coverage fell slightly below
nominal; the LASSO loss (deviance vs.~MSE) had negligible impact. This
mild undercoverage of the TMLE-based pipelines under rarity (LASSO-TMLE,
C-TMLE, and Standard TMLE, about 91--93\%) is consistent with
finite-sample targeting when events are sparse: with few informative
observations the efficient-influence-curve variance is estimated
slightly anti-conservatively and the targeting step is less stable, so
the model-based SE modestly understates the empirical sampling variation
(negative relative error). These influence-curve SEs condition on the
cross-validation--selected nuisance model, whereas the bootstrap-based
SEs (OAL, GLiDeR, HAL) resample the full pipeline and thus also
propagate selection uncertainty; the calibration and runtime comparisons
are therefore partly confounded by which error sources each SE captures.

OAL (IPTW) and GLiDeR were comparatively stable across scenarios, with
well-controlled bias (OAL slightly below the truth, GLiDeR slightly
above) and coverage close to 95\%---a clear improvement over the
traditional LASSO approaches, likely stemming from their outcome-aware
selection and joint modeling.

HAL (G-Computation) produced highly concentrated estimates---tiny
empirical SE, coverage near one, and very large relative error: with a
fixed data-generating process and large sample the fitted surfaces vary
little across replicates, while the bootstrap inflates the model-based
SE. Its near-zero bias and tiny empirical SE give it the smallest RMSE
of any pipeline (Table S2), so its shortfall is one of interval
calibration and compute cost, not point accuracy. Adding a TMLE step
preserved low bias while improving coverage and relative error.

For C-TMLE, collaborative undersmoothing did not yield meaningful gains
and was sometimes nominally worse than Standard TMLE---plausibly
because, in this many-proxy setting, the extra variables retained under
small penalties added more noise than signal once the strong baseline
covariates captured most confounding.

\subsection{Practical recommendations}\label{practical-recommendations}

Synthesizing calibration and cost under the null benchmark yields a
practical ordering (Table 5); this is guidance for calibration and
residual-bias control at the null, not general effect-estimation advice.
For calibrated inference with compute available, OAL (IPTW) and GLiDeR
are preferred; HAL (TMLE) attains nearly the same calibration in minutes
and is the best performance-per-compute choice. When compute is
constrained, Standard TMLE and LASSO-TMLE are fast defaults that
under-cover only mildly under rarity. LASSO-IPTW and HAL (G-Computation)
are not recommended---the former biased under rare exposure, the latter
over-covering with very large relative error at over 16 hours per
single-core analysis.

\begin{table}[H]
\centering
\caption{\label{tab:tab6-guidance}Practical guidance integrating three axes: calibration (95\% confidence-interval coverage), robustness to exposure/outcome rarity (Note column), and computational cost (single-core runtime from Table 4). The ``Not recommended'' pipelines fail on bias (LASSO-IPTW under rare exposure) or over-cover with very large relative error at high cost (HAL G-Computation).}
\centering
\fontsize{8}{10}\selectfont
\begin{tabular}[t]{>{\raggedright\arraybackslash}p{3.0cm}>{\raggedright\arraybackslash}p{2.9cm}>{\raggedright\arraybackslash}p{1.5cm}>{\raggedright\arraybackslash}p{1.7cm}>{\raggedright\arraybackslash}p{3.0cm}}
\toprule
\cellcolor[HTML]{E8ECF0}{Priority} & \cellcolor[HTML]{E8ECF0}{Pipeline(s)} & \cellcolor[HTML]{E8ECF0}{Calibration} & \cellcolor[HTML]{E8ECF0}{Runtime} & \cellcolor[HTML]{E8ECF0}{Note}\\
\midrule
Calibrated inference, compute available & OAL (IPTW), GLiDeR & 94-95\% & 2-8 h & stable across rarity; bootstrap cost\\
Best calibration per unit compute & HAL (TMLE) & 94-95\% & \textasciitilde{}7 min & near-best calibration, far cheaper\\
Fast default (compute-limited) & Standard TMLE, LASSO (TMLE) & 91-95\% & <6 s & mild undercoverage under rarity\\
Not recommended & LASSO (IPTW); HAL (G-Comp.) & over-covers & <1 s / >16 h & rare-exposure bias; over-covering intervals\\
\bottomrule
\end{tabular}
\end{table}

\subsection{Limitations and future
directions}\label{limitations-and-future-directions}

Several limitations apply. First, HAL was fit with
\texttt{max\_degree\ =\ 1}---an additive first-order basis of indicator
(spline) terms in each covariate, without interactions---because higher
degrees were computationally infeasible at this scale (\(n = 4{,}500\)
with 167 candidate predictors). This still yields a flexible
nonparametric additive fit, and, reassuringly, the doubly robust HAL
(TMLE) pipeline remained among the best-calibrated methods despite it,
indicating the targeting step largely compensates for the restricted
basis. The cost of \texttt{max\_degree\ =\ 1} falls mainly on HAL
(G-Computation)---it helps explain that pipeline's over-concentrated
estimates---so allowing higher-order interactions might reduce that
over-concentration but is unlikely to change the doubly robust
conclusions materially. Second, the traditional LASSO pipelines enforced
inclusion of all investigator-specified covariates whereas the other
methods did not, a difference that may affect direct comparability.
Relatedly, the prescription-derived proxies are predominantly
\emph{outcome} predictors with weak treatment association, so the
simulation chiefly exercises outcome-aware selection rather than strong
treatment-side confounding; behavior may differ where proxies are
strongly treatment-associated. Third, the evaluation rests on a single
plasmode anchor under a null effect with a binary outcome and the risk
difference as the sole estimand, limiting generalizability to other data
structures, non-null effects, and estimands; all reported bias and
calibration therefore characterize behavior at the null rather than
power or bias at a true effect. Moreover, exposure is generated from a
correctly specified main-terms logistic propensity model on \(L\), so
the true propensity score lies within the parametric class fitted by the
PS-based pipelines (LASSO-IPTW, OAL, C-TMLE, Standard TMLE, GLiDeR);
this removes propensity-model misspecification as a bias source. Because
every PS-based pipeline shares this advantage equally it is common-mode
and does not explain the headline gaps---the rare-exposure differences
are driven by weighting instability rather than PS-model fit, and the
outcome-modeling and doubly robust arms do not depend on it---but the
near-nominal calibration of OAL and GLiDeR should still be read as
conditional on correct propensity specification. The design also fixes a
single static exposure with no incident-user time-zero, so it does not
address the time-varying confounding central to many longitudinal claims
studies; extending this comparison to new-user, active-comparator
designs is an important next step. Finally, more advanced hybrids---the
generalized OAL (GOAL)\textsuperscript{31} and the outcome HAL
(OHAL)\textsuperscript{13,14}---were prototyped but excluded by design.
We frame this study as a reproducible platform: each deferred
extension---non-null effect sizes, a sparse-signal higher-dimensional
regime, additional estimands and time-varying designs, and per-core
runtimes---is largely a one-argument change to the released
data-generating pipeline, which we make available for others to run.

\subsection{Conclusions}\label{conclusions}

Methods incorporating more targeted variable selection often
outperformed the baseline combination of LASSO and IPTW, yielding
better-calibrated intervals and smaller residual bias, and pairing these
selection approaches with a doubly robust estimator generally improved
performance---suggesting that flexible nuisance estimation combined with
targeting is effective with large proxy sets. The degree of improvement
varied across methods and scenarios, so estimator choice should be
guided by the performance metrics prioritized in a given application,
the rarity of the exposure and outcome, and the available computational
budget.

\section*{Ethics Statement}\label{ethics-statement}
\addcontentsline{toc}{section}{Ethics Statement}

This study is a secondary analysis of de-identified, publicly available
data from the National Health and Nutrition Examination Survey (NHANES)
2013--2018, collected by the National Center for Health Statistics
(NCHS). The NHANES protocol was approved by the NCHS Research Ethics
Review Board, and written informed consent was obtained from all
participants at the original data collection. Because the data are
publicly accessible, fully de-identified, and contain no personally
identifiable information, this secondary analysis did not constitute
human-subjects research requiring additional institutional review board
approval or separate participant consent. No new data were collected.
The study was conducted in accordance with the Guidelines for Good
Pharmacoepidemiology Practice (GPP); the simulation component is
reported following the ADEMP framework\textsuperscript{25}, and the
real-data motivating analysis---an illustrative cross-sectional example
rather than a benchmark---is reported following the STROBE statement for
observational studies.

\section*{Conflict of Interest}\label{conflict-of-interest}
\addcontentsline{toc}{section}{Conflict of Interest}

The authors declare no conflicts of interest.

\section*{Acknowledgements}\label{acknowledgements}
\addcontentsline{toc}{section}{Acknowledgements}

\textbf{Funding.} This research received no specific grant from any
funding agency in the public, commercial, or not-for-profit sectors.

\textbf{Use of artificial intelligence.} During this work, the authors
used AI-based tools (large language models) to assist with the analysis
and simulation code, and text editing; the authors verified all outputs
and take full responsibility for the content.

\section*{Transparency and Reproducibility
Statement}\label{transparency-and-reproducibility-statement}
\addcontentsline{toc}{section}{Transparency and Reproducibility
Statement}

Following the structure recommended by Wang and Pottegård (Am J
Epidemiol 2023), we report: (1) \emph{Protocol} --- no separately
registered protocol was prepared; the full simulation design and
analysis plan are reported in the Methods following the ADEMP framework.
(2) \emph{Preregistration} --- the study was not preregistered. (3)
\emph{Data access} --- the NHANES 2013--2018 data are publicly
available\textsuperscript{4}. The processed anchor cohort and the full
plasmode data-generation pipeline are bundled in the public code
repository (below); the large replicate datasets are not committed but
are regenerated from the bundled anchor by that pipeline. (4) \emph{Code
sharing} --- all code---the data-generation pipeline and single anchor
dataset, the method implementations, per-scenario runners, and the
figure generators---is publicly available in the self-contained
repository \url{https://github.com/ehsanx/hD-causal-comp-code} (built on
the \texttt{Plasmode} package\textsuperscript{23}). (5) \emph{Reporting
guidelines} --- the simulation follows ADEMP\textsuperscript{25} and the
real-data motivating analysis follows the STROBE statement.

\section*{References}\label{references}
\addcontentsline{toc}{section}{References}

\protect\phantomsection\label{refs}

\end{document}